\documentclass[preprint,prX]{revtex4}

\usepackage{graphicx,psfrag}
\usepackage[latin1]{inputenc}
\usepackage{natbib}
\usepackage{amsmath, amsthm, amssymb}
\usepackage{comment}
\usepackage{amsmath}

\newcommand{\ds}{\displaystyle}
\newcommand{\mb}{\mathbf}
\newcommand{\notes}[1]{}

\newcommand{\beq}{\begin{equation}}
\newcommand{\eeq}{\end{equation}}
\newcommand{\beqnn}{\begin{equation*}}
\newcommand{\eeqnn}{\end{equation*}}
\newcommand{\beqas}{\begin{eqnarray*}}
\newcommand{\eeqas}{\end{eqnarray*}}
\newcommand{\beqa}{\begin{eqnarray}}
\newcommand{\eeqa}{\end{eqnarray}}
\newcommand{\p}{\partial}

\begin{document}

\title{Stable Magnetic Droplet Solitons in Spin Transfer Nanocontacts}

\author{Ferran Maci\`a}
\affiliation{Grup de Magnetisme, Dept. de F\'isica, Universitat de Barcelona, Spain}
\affiliation{Department of Physics, New York University, New York, New York 10003, USA}
\author{Dirk Backes}
\affiliation{Department of Physics, New York University, New York, New York 10003, USA}
\author{Andrew D. Kent}
\affiliation{Department of Physics, New York University, New York, New York 10003, USA}
\date{\today}


\begin{abstract}

Magnetic thin films with perpendicular magnetic anisotropy (PMA) have localized excitations that correspond to reversed dynamically precessing magnetic moments, known as magnetic droplet solitons. Fundamentally, these excitations are associated with an attractive interaction between elementary spin-excitations (i.e., magnons) and were predicted to occur in PMA materials in the absence of damping \cite{Ivanov1977,Kosevich1990}. While damping, present in all magnetic materials, suppresses these excitations, it is now possible to compensate damping by spin transfer torques through electrical current flow in nanometer scale contacts to ferromagnetic thin films \cite{Slonczewski1996,Slonczewski2}. A theory predicts the appearance of magnetic droplet solitons at a threshold current in nanocontacts \cite{Hoefer2010} and, recently, experimental signatures of droplet nucleation have been reported \cite{scienceDroplet2013}. However, thus far, they have been observed to be nearly reversible excitations, with only partially reversed magnetization and to be subject to instabilities that cause them to drift away from the nanocontacts (i.e., drift instabilities) \cite{scienceDroplet2013}. Here we show that magnetic droplet solitons can be stabilized in a spin transfer nanocontact. Further, they exhibit a strong hysteretic response to fields and currents and a nearly fully reversed magnetization in the contact. These observations, in addition to their fundamental interest, open up new applications for magnetic droplet solitons as multi-state high frequency current and field tunable oscillators.
\end{abstract}

\maketitle

Spin transfer torque nanooscillators (STNO) are nanometer scale electrical contacts to ferromagnetic thin films that consist of a free magnetic layer (FL) and a fixed spin-polarizing magnetic layer \cite{Tsoi,Kiselev,Rippard2004,Bonettiprl2010}. The spin-transfer torque in such contacts can compensate the damping torque and excite spin-waves in the free layer, at a threshold dc current. When these spin-waves have a frequency less than the lowest propagating spin-wave modes in the ferromagnetic film, the ferromagnetic resonance (FMR) frequency \cite{kittel1}, they are localized in the contact region. In PMA free layers this has been predicted to lead to \emph{dissipative droplet solitons} \cite{Hoefer2010} (hereafter referred to as droplet solitons), which are related to the conservative magnon droplets that were studied in uniaxial (easy axis type) ferromagnets with no damping \cite{Ivanov1977,Kosevich1990}. In the nanocontact, energy dissipated due to damping (essentially friction) is compensated by energy input associated with spin-transfer torques, resulting in steady state spin-precession. Droplet solitons are expected to be strongly localized in the contact region, as well as to have spins precessing in-phase in the film plane  \cite{Hoefer2010}. In addition, for sufficient current the magnetization in the contact was predicted to be almost completely reversed relative to the film magnetization outside the contact. While precession frequencies below the FMR frequency were observed to appear at a threshold current in recent experiments \cite{scienceDroplet2013}, there was no evidence for fully reversed spins in the contact. Here we report evidence for nearly full reversal of the magnetization in the contact region. Further, we show that droplet solitons are stable, and exhibit a strong hysteretic response to currents and applied fields. We also observe structure in their resistance versus field characteristics, suggesting they experience a disordered pinning potential. These results are important to understanding and controlling their motion, nucleation and annihilation \cite{Hoefer2012, Hoefer2014}.

\begin{figure*}[t]
\includegraphics[width=\columnwidth]{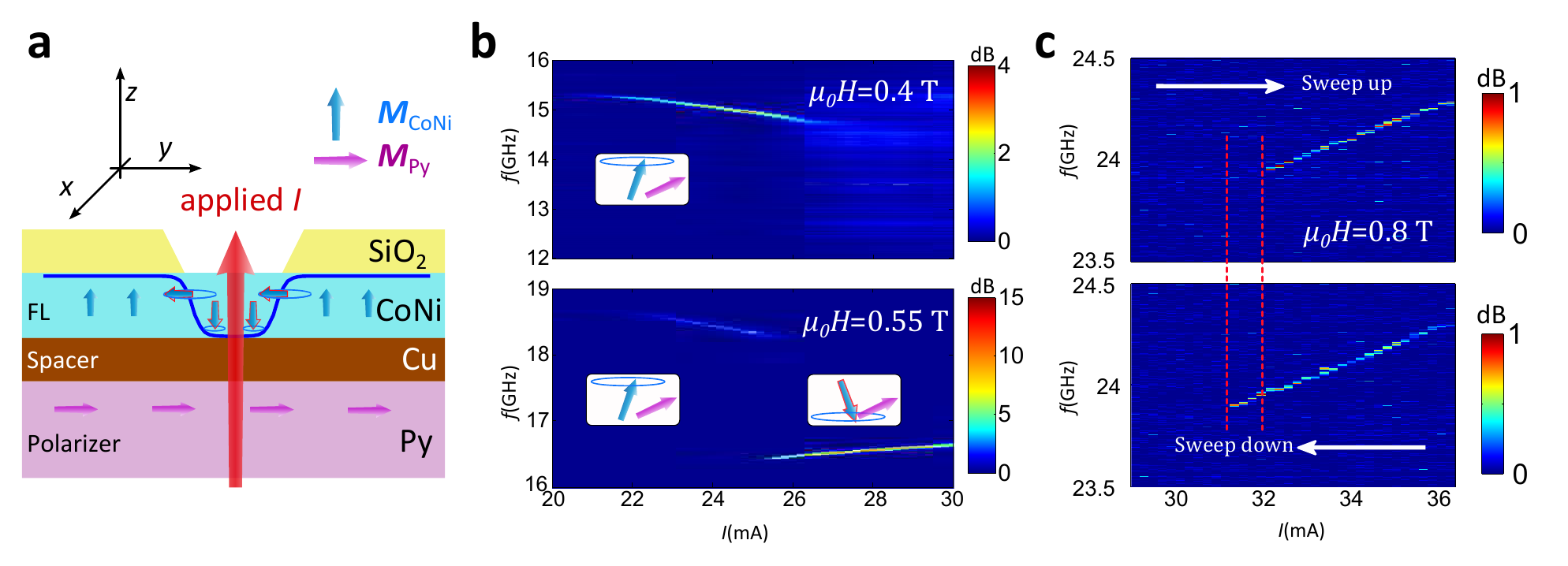}
\caption{\small{\textbf{Device schematic and dynamical properties.} (a) Schematic of a magnetic droplet in a STNO. An electrical current flows through a nanocontact to a thin ferromagnetic layer (the free layer, FL) towards a fixed spin-polarizing layer. The external magnetic field is applied perpendicular to the film plane (the $z$-direction in the figure). The droplet is a nearly reversed magnetization region with spins precessing in the $x-y$ plane. (b \& c) High frequency spectra as a function of the current in applied fields of (b) $\mu_0 H=0.4$ and $0.55$ T and (c) $\mu_0 H=0.8$ T, showing both field swept-up and swept-down measurements.  Hysteresis is found, as indicated by the vertical dashed lines.}}
\label{fig1}
\end{figure*}

To study droplet solitons we fabricated STNO with a free layer with PMA and an in-plane magnetized polarizing layer (shown schematically in Fig.~\ref{fig1}a) to measure their dc and high frequency electrical characteristics. As the electrical signal in our STNO is associated with the giant magnetoresistance effect, the in-plane magnetized polarizer allows us to detect in-plane precession of the magnetization of the free layer in the contact region, as discussed further below. Our layer structure consisted of a $4$ nm thick  Cobalt-Nickel (CoNi) multilayer, $10$ nm of Copper (Cu) followed by $10$ nm of Permalloy (Ni$_{80}$Fe$_{20}$, denoted Py) \cite{Macia:jmmm2012}.  The CoNi multilayer is the free layer and has an easy magnetization axis perpendicular to the plane, while the Py layer serves as a polarizer and has its magnetization in the film plane. The Cu layer decouples the magnetic layers while enabling spin-transport between the layers (i.e., the Cu layer is much thinner than its spin-diffusion length).  We defined contacts to the films that ranged from $70$ to $200$ nm in nominal diameter. Further, our sample layout enables electronic transport measurements from dc to microwave frequencies ($\sim$50 GHz).

We characterized the magnetic properties of the layer stacks using ferromagnetic resonance (FMR) spectroscopy. Most important for these studies, is that the CoNi has an effective perpendicular anisotropy field of $\mu_0 H_P=0.25$ T, which is defined to be the perpendicular anisotropy field minus the saturation magnetization, $H_P=H_K-M_s$. We also studied samples with smaller $\mu_0H_P \simeq 0.1$ T. We focus on the results for the STNO with $\mu_0 H_P=0.25$ T in this paper. (Results for the other series of STNO are presented in the Supplementary Materials section.)

As noted, the electrical response of our samples is associated with the giant magnetoresistance effect. The device resistance depends on the relative magnetization alignment of the free and polarizing layers, $\mathrm{MR}=(R(H)-R_P)/R_P=R_0(1-\mathbf{\hat{m}}_\mathrm{FL} \cdot \mathbf{\hat{m}}_\mathrm{P})/2$, where $\mathbf{\hat{m}}_\mathrm{FL/P}$ are unit vectors in the direction of the FL and polarizer magnetization respectively and $R_0=(R_\mathrm{AP}-R_\mathrm{P})/R_P$ is the fractional change in resistance between the device antiparallel (AP) and parallel (P) magnetization states. A magnetic field applied perpendicular to the film plane tilts the Py magnetic moments out of the film plane, $m_{z,\mathrm{P}}=H/M_s$ for $H<M_s$ with $\mu_0 M_s\simeq 1.1$ T, while the CoNi magnetic moments remain perpendicular to the film plane, $m_{z,\mathrm{FL}}=1$. Thus, for $\mu_0 H> \mu_0 M_s = 1.1$ T, the layer magnetizations align forming a P state. This results in the resistance of the STNO decreasing linearly with the applied field for $H<M_s$ and saturating when $H>M_s$. On the other hand, precession of the in-plane component of the FL magnetization leads to an oscillating resistance and thus a voltage response in the microwave range for fields less than the saturation field of the polarizing layer (i.e., when the polarizing layer has a component of magnetization in the film plane).

Figure~\ref{fig1}b shows measurements of the STNO high frequency response versus dc current at two different applied perpendicular magnetic fields at room temperature. At $0.4$ T ($m_{z,\mathrm{P}}\simeq0.36$) the STNO signal output frequency decreases with increasing bias current (i.e., there is a \emph{redshift} of the signal) \cite{Rippard2010,akerman_pss2011}. While at $0.55$ T ($m_{z,\mathrm{P}}\simeq0.5$) there is initially a redshift of the oscillation frequency with increasing current, followed by an abrupt ($\sim 3$ GHz) decrease of the signal frequency at a threshold current--that has been associated with the creation of a droplet soliton \cite{scienceDroplet2013}. The signal frequency after the jump is close to the Zeeman frequency, the prediction for a reversed droplet soliton, $f_\mathrm{Zeeman}=\gamma \mu_0 H/(2\pi)$, where $\gamma$ is the gyromagnetic ratio \cite{Ivanov1977,Kosevich1990}. Figure~\ref{fig1}c shows that at $0.8$ T  ($m_{z,\mathrm{P}}\simeq0.73$) we only observe the lower frequency (magnetic droplet) excitation and the frequency increases (blueshifts) with increasing current (Fig.~\ref{fig1}c). Most interestingly, at $0.8$ T the spectra are hysteretic$-$the spectra depend on the field sweep direction. The droplet soliton nucleates at about $32$ mA with increasing current and is annihilated at about $31$ mA with decreasing current. We note also that with increasing field the magnetization of the Py saturates perpendicular to the film plane$-$in the same direction as the CoNi$-$and thus the microwave signal vanishes. However, dc-magnetoresistance measurements still enable characterization of the droplet, because precession of FL changes its perpendicular component ($m_{z,\mathrm{FL}}$) and this results in a variation of the dc resistance (even when $m_{z,\mathrm{P}}=1$).

\begin{figure}[t]
\includegraphics[width=\columnwidth]{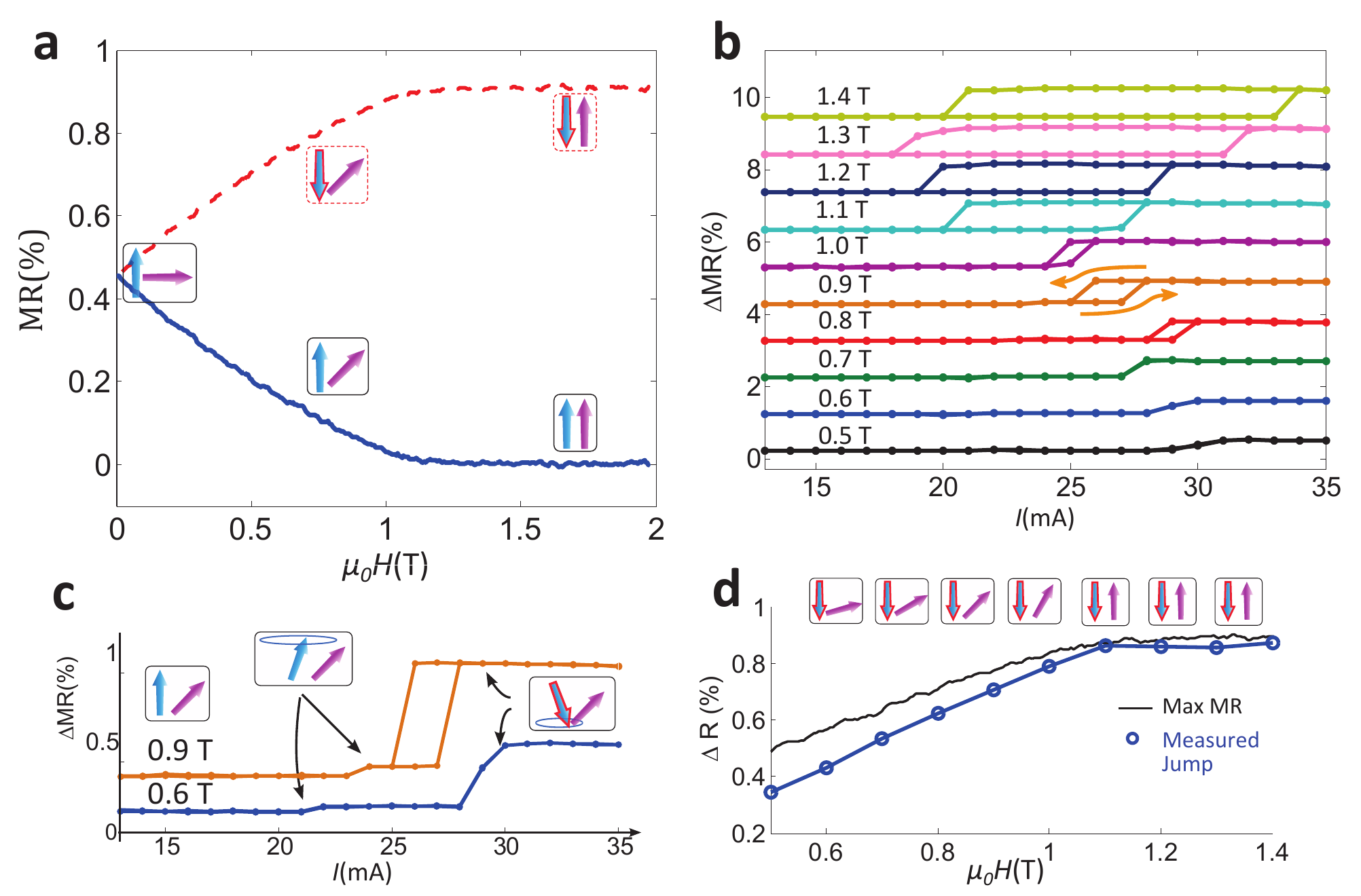}
\caption{\small{ \textbf{Nucleation of droplet solitons with current.} (a) Magnetoresistance of a $150$ nm diameter STNO versus field with $I=5$ mA. Light blue and pink arrows represent the magnetic moments for CoNi and Py; we outline the CoNi arrows that represent droplet soliton states. (b) MR as a function of the applied current for fields ranging from $\mu_0 H=0.5$ to 1.4 T. Curves are shifted vertically for clarity. (c) Expanded MR curves in (b) for the fields $\mu_0 H=0.6$ and 0.9 T, with the onset of spin-wave excitations and the droplet state indicated. (d) The MR in the droplet soliton state as a function of the applied field. The black curve is the expected maximum GMR of the STNO as a function of the applied field, $\mathrm{MR}=R_0 H/M_s$ where $R_0$ is 0.9\% and $\mu_0 M_s$ is 1.1 T.}}
\label{fig2}
\end{figure}

Hence, to characterize magnetic droplet excitations$-$both their onset and annihilation$-$we measured the dc magnetoresistance (MR) of our devices at low temperature ($4.2$ K) within a vector superconducting magnet. In Fig.~\ref{fig2}a we show the MR for fields applied perpendicular to the film plane; a resistance of zero corresponds to the fields at which the magnetizations of the Py and CoNi align parallel, and the overall MR of the STNO ($R_0=0.9$ \%) corresponds to twice the value at zero field when the magnetizations of the Py and CoNi are orthogonal. The dashed red curve in Fig.~\ref{fig2}a illustrates the expected MR for a reversed CoNi magnetization (i.e., magnetization antialigned with the applied field). This curve is obtained by reflecting the measured MR about the horizontal line, MR($H=0$). Fig.~\ref{fig2}b shows  current swept MR measurements at a series of perpendicularly applied magnetic fields. The signal to noise ratio increases at $4.2$ K and MR data can now be used to detect both the onset of spin-waves excitations, corresponding to a small $0.02-0.08$\% increase in MR, and the onset of droplet solitons, 0.85\% change in resistance (see, for instance, the curves in Fig.~\ref{fig2}c).

The resistance curves at constant applied fields (Fig.~\ref{fig2}b and c) show the onset of droplet solitons when increasing the current and the annihilation of the soliton states when decreasing the  current. There is hysteresis showing that there is an energy barrier separating STNO states and indicating that the droplet soliton is stable for a range of applied currents. In Fig.~\ref{fig2}c we see that at low fields ($\mu_0H \sim 0.6$ and 0.9 T) there is first a small step in resistance that corresponds to the onset of a (small angle precession) spin-wave excitation and next a larger step in resistance that corresponds to the nucleation of the droplet soliton. We have plotted the maximum change in the resistance step corresponding to the soliton excitation in Fig.\ \ref{fig2}d and compared it with the maximum expected change in resistance (i.e., $\mathrm{MR}=R_0 H/M_s$). The overall change is almost the full MR, indicating the CoNi magnetization is reversed nearly completely in the nanocontact area. The overall droplet MR increases with applied field because the relative alignment between Py and CoNi magnetization increases as the Py magnetization tilts with increasing applied field (see the schematic blue and pink arrows in Fig.\ \ref{fig2}d). The difference in resistance between that measured in the droplet state and full magnetization reversal of the FL ranges from 20\% at $H=0.6$ T to just 5\% at fields above $1$ T; this difference may be due to precession in the droplet that decreases the overall perpendicular component of the magnetization or by a small displacement of the droplet from the contact center.

\begin{figure}[t]
\includegraphics[width=\columnwidth]{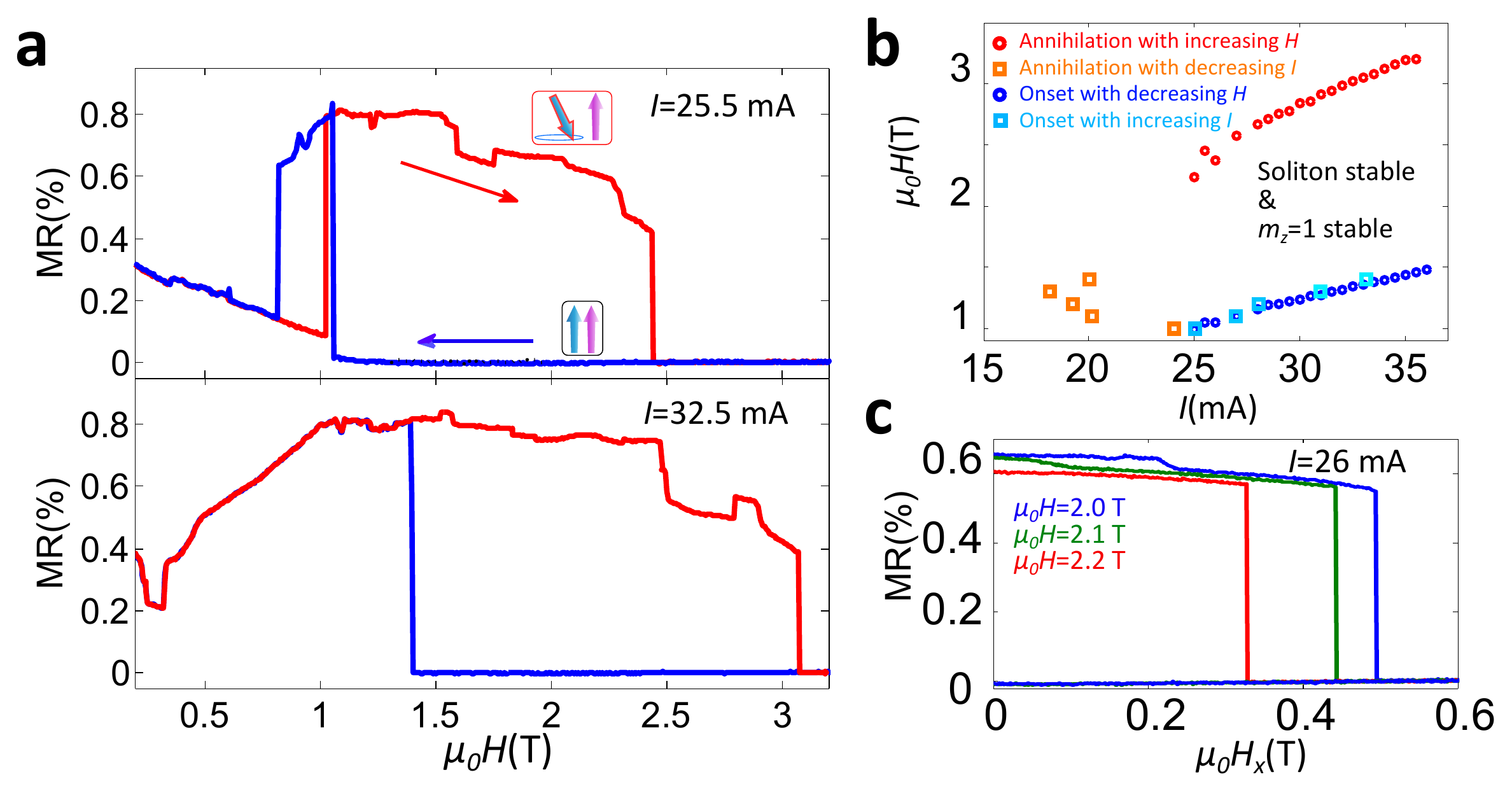}
\caption{\small{\textbf{Hysteresis in the magnetoresistance (MR) data.} (a) MR curves as a function of the perpendicular applied field for currents of $I=25.5$ mA and $I=32.5$ mA. (b) Stability map of droplet solitons: red circles show annihilation with increasing $H$, orange squares show annihilation with decreasing $I$, blue circles show onset with decreasing $H$, and cyan squares show onset with increasing $I$. (c) MR curves as a function of the in-plane field $H_x$ for an current of $I=26$ mA. The three curves correspond to a perpendicular applied field of 2.0 (blue) 2.1 (green), and 2.2 (red) T.}}
\label{fig3}
\end{figure}

Next  we analyze the onset of droplet solitons as a function of applied field at constant current. In Fig.~\ref{fig3}a we show the MR as a function of the applied field for two fixed applied currents, $I=25.5$ mA and $I=32.5$ mA. The MR curves clearly show both the onset and the annihilation of the droplet excitations as well as a remarkable hysteretic behavior, especially at large fields. The MR curve for $I=32.5$ mA (see Fig.\ \ref{fig3}a lower panel) shows the nucleation of a droplet soliton at a field of about $0.3$ T; at this field the magnetization of the CoNi layer reverses and opposes the applied field. The resistance then increases with the applied field until it saturates when the Py magnetization saturates (i.e. when the magnetizations of CoNi and Py are antiparallel). At even larger field ($\simeq 3$ T) there is a step decrease in resistance, which we associate with the droplet annihilation. When the applied field is reduced, the droplet nucleates at much lower field ($\simeq 1.4$ T). The large field hysteresis is consistent with the current swept data at fixed field, as we discuss further below.

We note that within the field range where droplet solitons are present there are additional and highly reproducible small steps in the resistance curves (Fig.~\ref{fig3}a). These may originate from pinning of the soliton at different sites within the contact. As the field increases, the droplet soliton state becomes less energetically favorable and the soliton might shift to different locations with slightly different effective fields caused by either current-induced Oersted fields or by small variations in the FL's magnetic anisotropy or magnetization. Such resistance states may also be due to changes in the droplet precession angles.

We next focus on the onset and annihilation of the soliton excitations when the Py layer is saturated ($\mu_0 H>1.1$ T) so the spin polarization of the current is constant with increasing field. There is hysteresis both in current sweeps at fixed field and field sweeps at fixed current. In Fig.~\ref{fig3}b we show the onset and annihilation conditions of the droplet soliton found for both cases. The points marking the onset of the soliton fall on a straight line (both for decreasing field and for increasing current). Annihilation occurs at larger fields and falls on a straight line as well--with a larger slope. The large area in between the two lines corresponds to the hysteretic region--the zone where the droplet soliton is present or absent depending on the STNO field and current history. Similar measurements on a sample with a smaller anisotropy ($H_P \sim 0.1$ T) also showed the straight lines for both the onset and annihilation of the droplet soliton state with a smaller area hysteretic region, a width in applied field of about $0.3$ T (see the Supplementary Materials).

We also observed that droplet solitons can be annihilated (and thus rendered unstable) with magnetic fields applied in the film plane. We nucleated droplet solitons at a large perpendicular magnetic field and then applied an in-plane field until we annihilate the droplet soliton. Figure\ \ref{fig3}c shows the MR as a function of the in-plane field at different perpendicular applied fields. We see that all the high-resistance states show an abrupt step down to the low-resistance state with increasing  in-plane field. Once we removed the in-plane field, we only nucleated droplet solitons in cases for which the perpendicular field made the soliton state stable and the no-soliton state unstable (i.e.,  perpendicular fields and currents that are not in the hysteretic zone in Fig.~\ref{fig3}b). Interestingly, the in-plane field values needed for droplet annihilation depend on what pinning state the soliton was in before applying the in-plane field. Again, we observed the effect in samples with smaller anisotropy.

We now consider the basic physics of the soliton nucleation and annihilation. Droplet solitons form in PMA thin films because spin torques can favor a layer magnetization antialigned with the magnetization of the polarizing layer (an AP state). Further, in PMA films ($H_P=H_k-M_s>0$) in a perpendicular applied field, any dynamical excitation tends to have a frequency that lies below the FMR frequency because the effective field ($\mathbf{H}_\mathrm{eff} = H_P m_z\mathbf{\hat{z}}$) decreases with increasing precession angle (i.e. $m_z<1$ and decreases with increasing precession angle).  Thus a soliton state is localized because its  excitation frequency is below the frequency of propagating spin waves modes. As the applied perpendicular field is increased the current required to sustain the droplet also increases and eventually the droplet annihilates. Hysteresis can result because the spin-torque required to maintain the droplet state is less than that required to nucleate it.

A quantitative understanding of droplet soliton hysteresis is possible through analysis of the Landau-Lifshitz-Gilbert-Slonczewski \cite{Slonczewski1996} equation describing the FL's magnetization dynamics. In dimensionless parameters the LLGS equation reads:
\beq
\frac{\p \mb{m}}{\p t}= \mb{m} \times \mb{h_\mathrm{eff}}-\alpha\mb{m} \times (\mb{m} \times \mb{h_{\text{eff}}})+\sigma_0\eta(m_z)\mb{m}\times (\mb{m} \times \mb{m_p}),
\label{lle}
\eeq
where the precession (first term) and damping (second term) include the effective field $\mb{h_{\text{eff}}}=\mb{h_0}+h_Pm_z\mathbf{\hat{z}}+\nabla^2 \mb{m}$, the sum of the applied field, the effective perpendicular anisotropy field, and the exchange field. The second term's coefficient $\alpha$ is the damping constant. The fields are normalized to the saturation magnetization, $M_s$ (e.g., $\mb{h_{0}}=\mb{H_{0}}/M_s$). The spin-torque (third term in Eq.\ \ref{lle}) includes the spin polarization direction of the applied current, $\mb{m_p}$, the torque asymmetry, $\eta(m_z)$ (defined below), and $\sigma_0$, which is proportional to the current intensity. (Further details on the analysis is in the Supplementary Materials.)

The simplest analysis considers a macrospin and thus does not include the exchange field or allow for spatial variations in the magnetization. The dynamical equation for the perpendicular component of magnetization in an applied field perpendicular, $\mb{h_0}=h_0\mathbf{\hat{z}}$, and with a polarization also perpendicular to the film's plane, $\mb{m_p}=\mathbf{\hat{z}}$, is
\beq
\dot{m}_z=-(1-m_z^2)(\sigma_0\eta(m_z)-\alpha (h_0+h_Pm_z)).
\label{mz}
\eeq
The state magnetized along the field direction, $m_z=1$, becomes unstable at a critical current and then it becomes stable again at a larger field. However,
the reversed magnetization, $m_z=-1$, has a different stability threshold, leading to hysteresis. For certain parameters there exists a third solution, $0<m_z<1$ but it is unstable for the form of the spin-torque and spin-torque asymmetry that we consider.

The stability thresholds for the two solutions gives a state diagram, the range of current and field parameters in which there is bistability. For $m_z=1$ the threshold is
\beq
h_0=\sigma_0\eta(m_z)/\alpha-(h_k-1)
\eeq
and for $m_z=-1$ it is:
\beq
h_0=\sigma_0\eta(m_z)/\alpha+(h_k-1).
\eeq
In the  case with no spin-torque asymmetry, $\eta=1$, the state diagram in applied field versus current ($h_0$ vs $\sigma_0$) is two straight lines separating $m_z=\pm 1$ transitions. The width of the hysteresis in applied field is twice the effective anisotropy field, $2h_P$. With an asymmetric spin torque \cite{Slonczewski1996}  $\eta=2\lambda^2/\left[(\lambda^2+1)+(\lambda^2-1)m_z\right]$ with $m_z=1$ we have the same stability condition but for $m_z=-1$ the threshold slope changes. Thus an asymmetric spin torque gives a field hysteresis that increases with the applied current, having a width in field of twice the effective perpendicular anisotropy field plus a term linear in the applied current, $2h_P+\sigma_0(\lambda^2-1)$.

The experimental data in Fig.~\ref{fig3}b shows two straight threshold lines with different slopes. Fitting this data to a macrospin model gives an asymmetric spin torque $\lambda^2=1.8$, with a hysteresis extrapolated to zero current of $0.7$ T, which is larger than twice the layer measured effective perpendicular anisotropy field ($0.25$ T). A state diagram for a STNO with an effective perpendicular anisotropy field of $\sim 0.1$ T was measured as well and could be fit assuming no spin-torque asymmetry ($\eta=1$) and also showed a smaller field hysteresis of $0.3$ T (see the Supplementary Materials).

We also considered a 2D model with exchange interactions and performed micromagnetic simulations. Specifically, we numerically solved the LLGS equation (Eq.\ \ref{lle}) to study the dynamics of the FL's magnetization. Again, we considered the case when the permalloy layer is saturated (i.e., the current polarization direction is constant, $\mb{m_p}=\mb{\hat{z}}$). We found that droplet solitons are created at a critical current \cite{Hoefer2010} and annihilate in a perpendicular applied field as well as with small in-plane magnetic fields. Our results show that the hysteresis is the same as that of the macrospin model. The main difference is that the lines separating the stability of the droplet soliton are shifted towards larger current densities--or towards lower applied fields--owing to the fact that the exchange induces diffusion in the system (see Supplementary Material).

Our simulations also show how droplet solitons annihilate with in-plane fields. In-plane fields cause the droplet soliton to delocalize and eventually lose stability; this occurs when the localized oscillations in the magnetization couple with the film's propagating spin-wave modes. Our analysis and experiments also show that droplet solitons are stable at relatively small perpendicular field fields. The macrospin model also predicts stable droplets at zero applied field, provided the polarizing layer is perpendicularly magnetized at zero field (i.e. has a net PMA). We also found this in micromagnetic simulations with a perpendicularly magnetized polarizer layer.

In summary we have demonstrated stable droplet solitons in nanocontacts to ferromagnetic thin films. Our experimental results reveal a nearly complete reversal of the magnetization in the nanocontact and a large hysteresis between the onset and annihilation of soliton excitations--both in field and current. This provides a new means for droplet solitons to carry and store information, in addition to in their phase and amplitude \cite{Hoefer2014}. Our modeling and micromagnetic simulations capture the main features of our experimental results and also predict that droplet solitons can exist at zero applied field. We have also observed small and reproducible variations on the nanocontact resistance when varying the field suggesting that the droplet solitons can be trapped at pinning sites or may have discrete precessional states. The hysteresis and the observed discrete states could be useful in in an application that requires a multistate oscillator.

\section*{Methods}

The layer stacks consist of Co and Ni, $6\times$(0.2Co$|$0.6Ni) capped with 0.2Co$|$5 Pt, separated by 10 Cu from a 10 Py layer, deposited by thermal and e-beam evaporation in an ultra high vacuum chamber (thicknesses in nanometers). The Cu spacer layer was chosen to magnetically decouple the in-plane magnetized Py layer from the out-of-plane magnetized Co$|$Ni multilayer; the Pt capping layer was used to further enhance the interface-induced perpendicular magnetic anisotropy of the Co$|$Ni multilayer. Layer stacks were deposited on oxides silicon wafers. The point contacts were defined by etching holes in a $50$ nm thick silicon dioxide layer deposited on top of the films. Electron-beam lithography was used to defined the point contacts with diameters ranging from 70 to 200 nm. Devices were patterned with a bottom electrode and a top electrode into structures suitable for both dc and microwave electrical measurements using optical lithography and etching techniques. Most of the nanofabrication process were carried out at the Center for Functional Nanomaterials at Brookhaven National Lab. The contacts were characterized by atomic force microscopy and scanning electron microscopy.

Our layer stacks have been characterized before and after patterning with FMR spectroscopy (see Supplementary Materials) in order to determine the magnetic anisotropy of both Py and CoNi. We used frequencies ranging from 1 to 40 GHz as a function of the applied field at room temperature and a coplanar waveguide (CPW) to create the microwave fields along with a network analyzer to record the absorption signal.

For the high frequency measurements our samples were contacted with picoprobes to a current source and to a spectrum analyzer that recorded the signals in the presence of applied magnetic fields. We used a broadband low noise 20 dB amplifier. The dc transport measurements were carried out by contacting devices with wire bonds. Low temperature transport measurements were conducted at 4.2 K in a three-dimensional vector superconducting magnet capable of producing bipolar fields up to 8 T.

\newpage
\begin{center}
\begin{Large}
Supplementary materials\\
Stable Magnetic Droplet Solitons in Spin Transfer Nanocontacts
\end{Large}
\end{center}

\section{Additional measurements}

We have measured samples with a similar layer stacks to the ones presented in the main manuscript but with a slightly different anisotropy field, $\mu_0 H_P$ ($=\mu_0(H_K-M_s)$), of about 0.1 T, determined using ferromagnetic resonance spectroscopy.  While the nominal thicknesses of the Co and Ni layers were the same, we obtained slightly different anisotropies in different deposition runs.

We have analyzed the stability of soliton excitations with applied magnetic fields similar to the method presented in Fig.\ 3 of the main manuscript.
\begin{figure}[b]
\includegraphics[width=\columnwidth]{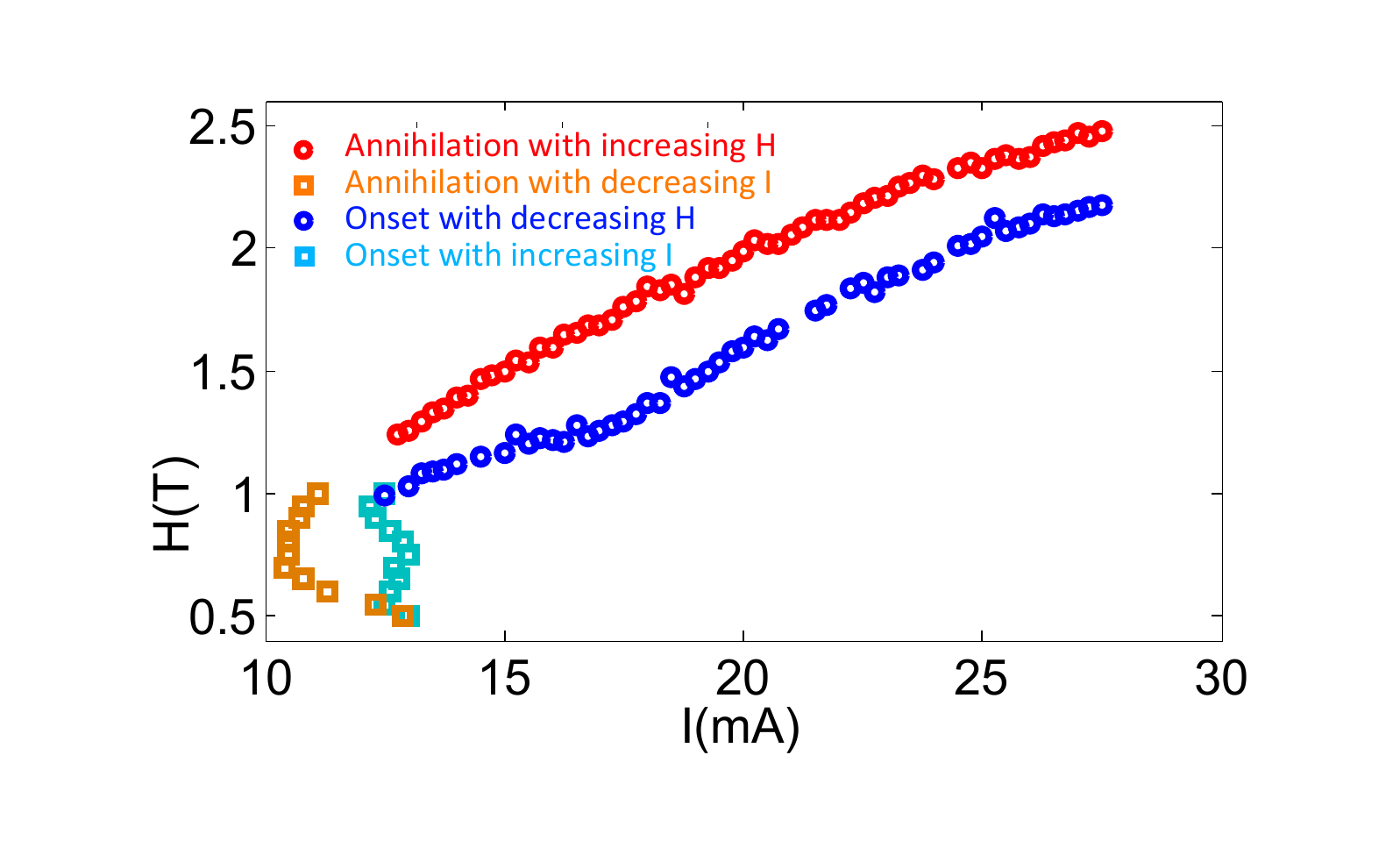}
\caption{\small{(a) Stability map of soliton excitations for a sample with an effective perpendicular anisotropy field $\mu_0 H_P$, of about 0.1 T: red circles show annihilation with increasing $H$, orange squares show annihilation with decreasing $I$, blue circles show onset with decreasing $H$, and cyan squares show onset with increasing $I$.}}
\label{fig1SM}
\end{figure}
Figure\ \ref{fig1SM} shows a stability map of soliton excitations as a function of the perpendicular applied field for a point contact with a nominal radius of $50$ nm. At fields larger than $1.1$ T when the Py layer is saturated, we can take the spin polarization of the current as a constant. The points separating the onset and annihilation of the droplet state fall on a nearly straight line and a hysteresis width in field of about $0.3$ T. We note here that this sample presents equal slopes for both onset and annihilation field versus current, suggesting the effect of the asymmetry in the spin torque is negligible (i.e. $\eta=1$).

At fields below 1.1 T, the onset current (and the annihilation current) decreases with increasing field ($\sim 1/H$) because the Py layer is not saturated and the polarization of the current and the spin-torque are mainly driven by the $z$-component of the magnetization of the polarizing Py layer.

\section{Modeling and Micromagnetics}

To understand the regions in applied field and current density where the droplet soliton state is stable we discuss in more detail the argument used in the main manuscript, both the macrospin model and micromagnetic simulations.

We again consider first the simplified macrospin model that neglects spatial variations in the free layer magnetization. We start with the LLGS equation for the magnetization dynamics:
\beq
\frac{\p \mb{m}}{\p t}= \mb{m} \times \mb{h_{eff}}-\alpha\mb{m} \times (\mb{m} \times \mb{h_{eff}})+\sigma\mb{m}\times (\mb{m} \times \mb{m_p}),
\label{lle}
\eeq
with $\mb{h_{eff}}=\mb{h_0}+h_Pm_z\mb{\hat{z}}$, $\alpha$ is the damping, and $\sigma$ is proportional to the current. Notice there is no exchange and $h_P>0$ is the effective perpendicular anisotropy, $h_P=h_k-1$. We have normalized magnetization and time by the saturation magnetization, $M_s$, and Larmor frequency, $\gamma \mu_0 M_s$, respectively.

The dynamics for the magnetization $z$ component is given by:
\beq
\dot{m}_z=-(1-m_z^2)(\sigma_0\eta(m_z)-\alpha h_{\text{eff}}),
\label{mz}
\eeq
where $\sigma=\sigma_0\eta$ and $\eta$ the polarization spin torque asymmetry. The effective field, $h_{\text{eff}}=h_0+h_P m_z$, includes the anisotropy, $h_P>0$, the external field in the $z$-direction, $h_0$, and the demagnetization field, $-m_z$.

We then perform a stability analysis of the $z$ component of the magnetization and see under what conditions the constant solutions, $m_z$, for Eq.\ \ref{mz}, are either stable ($\p_{m_z}(\p_tm_z)<0$) or unstable ($\p_{m_z}(\p_tm_z)>0$). This reasoning has already been used in \cite{mangin} and in \cite{Kent} to explain the switching of a nanopilar (in-plane and out-of-plane respectively).

Equation \ref{mz} can have either 2 or 3 solutions depending on the parameters $h_0$, $\sigma$, $\alpha$ and $h_k$:
\beq
\begin{array}{rcl}
m_z&=&\pm 1\\\\
m_z&=&\ds\frac{\sigma/\alpha-h_0}{h_k-1} ~~~\text{if}~ -1\le m_z\le1
\end{array}
\eeq
Here, we note that the third solution is always unstable if there is perpendicular anisotropy and for the form of the spin-torque torque and spin-torque asymmetry we consider. We consider a spin-transfer torque that depends on the angle between the free and polarizing layer's magnetization \cite{Slonczewski1996,stiles}:
\beq
\eta=\frac{2\lambda^2}{(\lambda^2+1)+(\lambda^2-1)m_z}.
\eeq
So $\lambda = 1$ is no asymmetry;  and $\lambda \ne 1$ results in $\eta(m_z = 1) = 1$ and $\eta(m_z = -1) = \lambda^2$.

One can derive the condition for a macrospin state to lose stability for the two solutions that correspond to the spins pointing along the applied field ($m_z = 1$) or in the opposite direction ($m_z = -1$).
\\
For $m_z=1$ we obtain
\beq
h_0=\sigma_0\eta(1)/\alpha-(h_k-1)
\eeq
and for $m_z=-1$ we have
\beq
h_0=\sigma_0\eta(-1)/\alpha+(h_k-1).
\eeq
We can now see that the hysteresis for a fixed current, $\sigma_0$, is:
\beq
\Delta h=\left(h_0+(h_k-1)\right)(\lambda^2-1)+2(h_k-1).
\eeq
\begin{figure}[htb]
\includegraphics[width=\columnwidth]{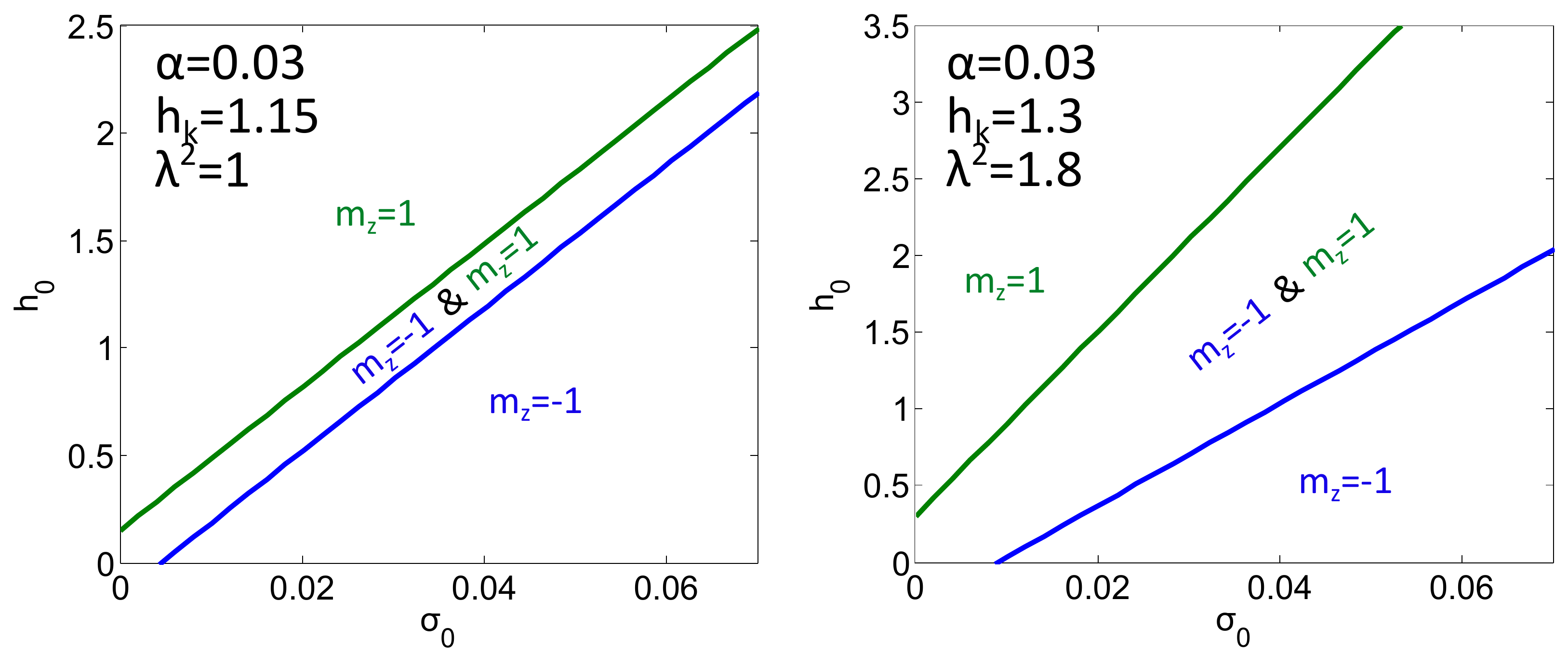}
\caption{\small{Stability thresholds versus applied current, $\sigma_0$, and field, $h_0$, separating stable and unstable regions for the solution $m_z=\pm1$. The damping value is taken at $\alpha=0.03$ and the anisotropy value, at $h_k=1.15$ and $h_k=1.3$ that corresponds to the values for our measured samples.}}
\label{fig2SM}
\end{figure}
In Fig.\ \ref{fig2SM} we plot the stability thresholds for $m_z=\pm 1$ using the parameters for CoNi multilayered film having a saturation magnetization value, $M_s$, of 0.96 T. We plotted two diagrams corresponding to: \emph{a)} an effective perpendicular anisotropy field $\mu_0(H_k-M_s)$ of 0.15 T and no spin torque asymmetry, $\lambda=1$, \emph{b)} an effective perpendicular anisotropy field $\mu_0(H_k-M_s)$ of 0.3 T and a spin torque asymmetry with $\lambda^2=1.8$.

This macrospin model has limitations because it neglects spatial variations in the magnetization.
We have done  2 dimensional micromagnetic simulations to investigate droplet excitations; onset, annihilation and hysteresis. We used the LLGS equation (Eq.\ \ref{lle}) with an effective field that includes exchange:
\beq
\mb{h_{\text{eff}}}=\mb{h_0}+(h_k-1)m_z\mb{z} + \nabla^2 \mb{m}.
\eeq

\begin{figure}[htb]
\includegraphics[width=\columnwidth]{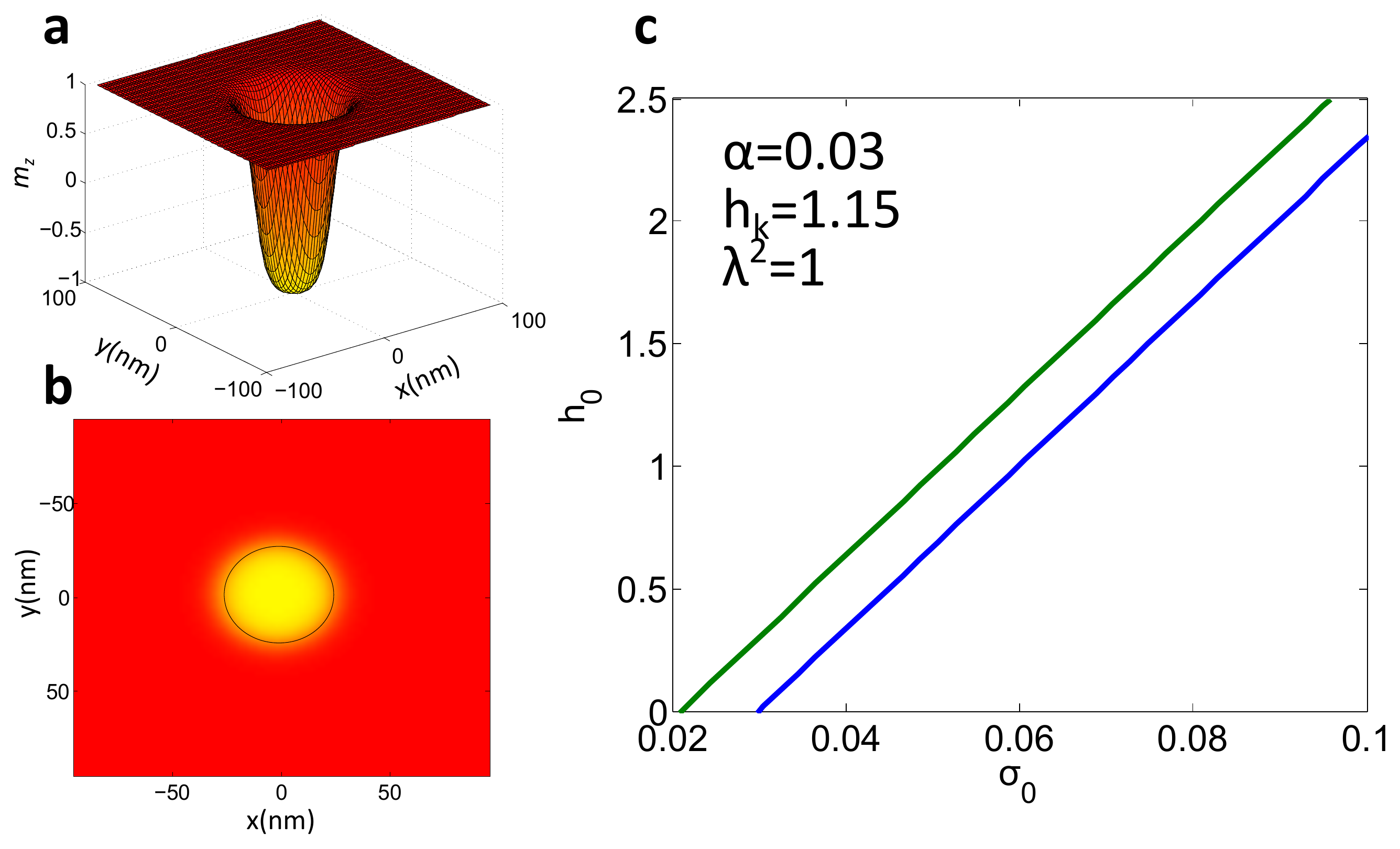}
\caption{\small{\textbf{Micromagnetic simulations} (a) and (b) Final state for the magnetization, $m_z$ in center of the nanocontact for a 50 nm point contact. The film parameters are $h_k=1.15$, $h_0=1.5$, $\sigma=0.1$ and $\alpha=0.03$. The simulation was run until a steady state was found corresponding to a reversed soliton. (c) Curves in applied current, $\sigma$, and field, $h$, separating stable and unstable regions for the reversed solitons . The damping value is taken at $\alpha=0.03$ and the anisotropy value, at $h_k=0.15$ }}
\label{fig3SM}
\end{figure}

We observe the formation of droplet solitons and their annihilation with both large out of plane fields and also with in-plane fields.
Figure\ \ref{fig3SM}a and b show the final state of a 50 nm point contact with an applied current, $\sigma_0$, of 0.1 under an applied field of $h_0=$1.5. We have considered a film with an effective field of 0.15 T ($h_k\approx 1.15$) and a damping, $\alpha$ of $0.03$. In this case it does not matter what the initial state is because the droplet is stable and it always forms no matter what the initial state is.
To study the hysteresis in the micromagnetic simulations we compared different applied fields, $h_0$, and current densities, $\sigma_0$, while keeping the parameters $h_k$ and $\alpha$ constant. In all cases we simulated 2 events, \emph{i)} initial condition is a nucleated droplet and \emph{ii)} initial condition is magnetization pointing in the applied field direction. We see hysteresis that has a size of twice the effective perpendicular anisotropy $h_k-1$ and shows a linear dependence between the applied current, $\sigma$ and the applied field, $h_0$, with a constant $\alpha$. In \cite{Hoefer2010} it was shown, with some analytical treatment of the LLGS equation (eq.\ \ref{lle}), the critical sustaining current depended linearly on the external $h_0$ with a factor $\alpha$.

We have computed the stability diagram showing the final state as a function of the initial conditions. Figure\ \ref{fig3SM}c corresponds to a film with $\mu_0(H_k-M_s)=0.15$ T of PMA and we neglected spin torque asymmetry.

This modeling shows that one should expect annihilation of the reversed excitations and that the hysteresis in the reversing is indeed intrinsic in the solution of the LLGS equation. It also tells that the size of the hysteresis is $2h_P$ (i.e., it increases with the anisotropy). The model also predicts that if we were able to maintain the polarization of the current at zero field we could create a droplet just increasing the current, which could be achieved with a perpendicular magnetized polarizer.
\\

\section*{Acknowledgements}
F.M. acknowledges support from EC, MC-IOF 253214, from Catalan Government COFUND-FP7, and from MAT2011-23698. This research was supported by NSF-DMR-1309202 and in part by ARO-MURI, Grant No. W911NF-08-1-0317. Research carried out in part at the Center for Functional Nanomaterials, Brookhaven National Laboratory, which is supported by the U.S. Department of Energy, Office of Basic Energy Sciences, under Contract No. DE-AC02-98CH10886.

\bibliographystyle{srt}

\begin{thebibliography}{19}
\expandafter\ifx\csname natexlab\endcsname\relax\def\natexlab#1{#1}\fi
\expandafter\ifx\csname bibnamefont\endcsname\relax
  \def\bibnamefont#1{#1}\fi
\expandafter\ifx\csname bibfnamefont\endcsname\relax
  \def\bibfnamefont#1{#1}\fi
\expandafter\ifx\csname citenamefont\endcsname\relax
  \def\citenamefont#1{#1}\fi
\expandafter\ifx\csname url\endcsname\relax
  \def\url#1{\texttt{#1}}\fi
\expandafter\ifx\csname urlprefix\endcsname\relax\def\urlprefix{URL }\fi
\providecommand{\bibinfo}[2]{#2}
\providecommand{\eprint}[2][]{\url{#2}}

\bibitem[{\citenamefont{Ivanov and Kosevich}(1977)}]{Ivanov1977}
\bibinfo{author}{\bibfnamefont{A.}~\bibnamefont{Ivanov}} \bibnamefont{and}
  \bibinfo{author}{\bibfnamefont{A.~M.} \bibnamefont{Kosevich}},
  \bibinfo{journal}{Zh. Eksp. Teor. Fiz.} \textbf{\bibinfo{volume}{72}},
  \bibinfo{pages}{2000} (\bibinfo{year}{1977}).

\bibitem[{\citenamefont{Kosevich et~al.}(1990)\citenamefont{Kosevich, Ivanov,
  and Kovalev}}]{Kosevich1990}
\bibinfo{author}{\bibfnamefont{A.~M.} \bibnamefont{Kosevich}},
  \bibinfo{author}{\bibfnamefont{B.~A.} \bibnamefont{Ivanov}},
  \bibnamefont{and} \bibinfo{author}{\bibfnamefont{A.~S.}
  \bibnamefont{Kovalev}}, \bibinfo{journal}{Phys. Rep.}
  \textbf{\bibinfo{volume}{194}}, \bibinfo{pages}{117} (\bibinfo{year}{1990}).

\bibitem[{\citenamefont{Slonczewski}(1996)}]{Slonczewski1996}
\bibinfo{author}{\bibfnamefont{J.~C.} \bibnamefont{Slonczewski}},
  \bibinfo{journal}{Journal of Magnetism and Magnetic Materials}
  \textbf{\bibinfo{volume}{1-2}}, \bibinfo{pages}{L1 } (\bibinfo{year}{1996}).

\bibitem[{\citenamefont{Slonczewski}(1999)}]{Slonczewski2}
\bibinfo{author}{\bibfnamefont{J.~C.} \bibnamefont{Slonczewski}},
  \bibinfo{journal}{Journal of Magnetism and Magnetic Materials}
  \textbf{\bibinfo{volume}{195}}, \bibinfo{pages}{L261} (\bibinfo{year}{1999}).

\bibitem[{\citenamefont{Hoefer et~al.}(2010)\citenamefont{Hoefer, Silva, and
  Keller}}]{Hoefer2010}
\bibinfo{author}{\bibfnamefont{M.~A.} \bibnamefont{Hoefer}},
  \bibinfo{author}{\bibfnamefont{T.~J.} \bibnamefont{Silva}}, \bibnamefont{and}
  \bibinfo{author}{\bibfnamefont{M.~W.} \bibnamefont{Keller}},
  \bibinfo{journal}{Phys. Rev. B} \textbf{\bibinfo{volume}{82}},
  \bibinfo{pages}{054432} (\bibinfo{year}{2010}).

\bibitem[{\citenamefont{Mohseni et~al.}(2013)\citenamefont{Mohseni, Sani,
  Persson, Anh~Nguyen, Chung, Pogoryelov, Muduli, Iacocca, Eklund, Dumas
  et~al.}}]{scienceDroplet2013}
\bibinfo{author}{\bibfnamefont{S.~M.} \bibnamefont{Mohseni}},
  \bibinfo{author}{\bibfnamefont{S.~R.} \bibnamefont{Sani}},
  \bibinfo{author}{\bibfnamefont{J.}~\bibnamefont{Persson}},
  \bibinfo{author}{\bibfnamefont{T.~N.} \bibnamefont{Anh~Nguyen}},
  \bibinfo{author}{\bibfnamefont{S.}~\bibnamefont{Chung}},
  \bibinfo{author}{\bibfnamefont{Y.}~\bibnamefont{Pogoryelov}},
  \bibinfo{author}{\bibfnamefont{P.~K.} \bibnamefont{Muduli}},
  \bibinfo{author}{\bibfnamefont{E.}~\bibnamefont{Iacocca}},
  \bibinfo{author}{\bibfnamefont{A.}~\bibnamefont{Eklund}},
  \bibinfo{author}{\bibfnamefont{R.~K.} \bibnamefont{Dumas}},
  \bibnamefont{et~al.}, \bibinfo{journal}{Science}
  \textbf{\bibinfo{volume}{339}}, \bibinfo{pages}{1295} (\bibinfo{year}{2013}).

\bibitem[{\citenamefont{Tsoi et~al.}(2000)\citenamefont{Tsoi, Jansen, Bass,
  Chiang, V., and Wyder}}]{Tsoi}
\bibinfo{author}{\bibfnamefont{M.}~\bibnamefont{Tsoi}},
  \bibinfo{author}{\bibfnamefont{A.~G.~M.} \bibnamefont{Jansen}},
  \bibinfo{author}{\bibfnamefont{J.}~\bibnamefont{Bass}},
  \bibinfo{author}{\bibfnamefont{W.-C.} \bibnamefont{Chiang}},
  \bibinfo{author}{\bibfnamefont{T.}~\bibnamefont{V.}}, \bibnamefont{and}
  \bibinfo{author}{\bibfnamefont{P.}~\bibnamefont{Wyder}},
  \bibinfo{journal}{Nature} \textbf{\bibinfo{volume}{406}}, \bibinfo{pages}{46}
  (\bibinfo{year}{2000}).

\bibitem[{\citenamefont{Kiselev et~al.}(2003)\citenamefont{Kiselev, Sankey,
  Krivorotov, Emley, Schoelkopf, Buhrman, and Ralph}}]{Kiselev}
\bibinfo{author}{\bibfnamefont{S.~I.} \bibnamefont{Kiselev}},
  \bibinfo{author}{\bibfnamefont{J.~C.} \bibnamefont{Sankey}},
  \bibinfo{author}{\bibfnamefont{I.~N.} \bibnamefont{Krivorotov}},
  \bibinfo{author}{\bibfnamefont{N.~C.} \bibnamefont{Emley}},
  \bibinfo{author}{\bibfnamefont{R.~J.} \bibnamefont{Schoelkopf}},
  \bibinfo{author}{\bibfnamefont{R.~A.} \bibnamefont{Buhrman}},
  \bibnamefont{and} \bibinfo{author}{\bibfnamefont{D.~C.} \bibnamefont{Ralph}},
  \bibinfo{journal}{Nature} \textbf{\bibinfo{volume}{425}},
  \bibinfo{pages}{380} (\bibinfo{year}{2003}).

\bibitem[{\citenamefont{Rippard et~al.}(2004)\citenamefont{Rippard, Pufall,
  Kaka, and Silva}}]{Rippard2004}
\bibinfo{author}{\bibfnamefont{W.}~\bibnamefont{Rippard}},
  \bibinfo{author}{\bibfnamefont{M.}~\bibnamefont{Pufall}},
  \bibinfo{author}{\bibfnamefont{S.}~\bibnamefont{Kaka},
  \bibfnamefont{S.~Russek}}, \bibnamefont{and}
  \bibinfo{author}{\bibfnamefont{T.}~\bibnamefont{Silva}},
  \bibinfo{journal}{Phys. Rev. Lett} \textbf{\bibinfo{volume}{92}},
  \bibinfo{pages}{027201} (\bibinfo{year}{2004}).

\bibitem[{\citenamefont{Bonetti et~al.}(2010)\citenamefont{Bonetti,
  Tiberkevich, Consolo, Finocchio, Muduli, Mancoff, Slavin, and
  \AA{}kerman}}]{Bonettiprl2010}
\bibinfo{author}{\bibfnamefont{S.}~\bibnamefont{Bonetti}},
  \bibinfo{author}{\bibfnamefont{V.}~\bibnamefont{Tiberkevich}},
  \bibinfo{author}{\bibfnamefont{G.}~\bibnamefont{Consolo}},
  \bibinfo{author}{\bibfnamefont{G.}~\bibnamefont{Finocchio}},
  \bibinfo{author}{\bibfnamefont{P.}~\bibnamefont{Muduli}},
  \bibinfo{author}{\bibfnamefont{F.}~\bibnamefont{Mancoff}},
  \bibinfo{author}{\bibfnamefont{A.}~\bibnamefont{Slavin}}, \bibnamefont{and}
  \bibinfo{author}{\bibfnamefont{J.}~\bibnamefont{\AA{}kerman}},
  \bibinfo{journal}{Phys. Rev. Lett.} \textbf{\bibinfo{volume}{105}},
  \bibinfo{pages}{217204} (\bibinfo{year}{2010}).

\bibitem[{\citenamefont{Kittel}(1946)}]{kittel1}
\bibinfo{author}{\bibfnamefont{C.}~\bibnamefont{Kittel}},
  \bibinfo{journal}{Phys. Rev.} \textbf{\bibinfo{volume}{70}},
  \bibinfo{pages}{965} (\bibinfo{year}{1946}).

\bibitem[{\citenamefont{Hoefer et~al.}(2012)\citenamefont{Hoefer, Sommacal, and
  Silva}}]{Hoefer2012}
\bibinfo{author}{\bibfnamefont{M.~A.} \bibnamefont{Hoefer}},
  \bibinfo{author}{\bibfnamefont{M.}~\bibnamefont{Sommacal}}, \bibnamefont{and}
  \bibinfo{author}{\bibfnamefont{T.~J.} \bibnamefont{Silva}},
  \bibinfo{journal}{Phys. Rev. B} \textbf{\bibinfo{volume}{85}},
  \bibinfo{pages}{214433} (\bibinfo{year}{2012}).

\bibitem[{\citenamefont{Maiden et~al.}(2014)\citenamefont{Maiden, Bookman, and
  Hoefer}}]{Hoefer2014}
\bibinfo{author}{\bibfnamefont{M.~D.} \bibnamefont{Maiden}},
  \bibinfo{author}{\bibfnamefont{L.~D.} \bibnamefont{Bookman}},
  \bibnamefont{and} \bibinfo{author}{\bibfnamefont{M.~A.}
  \bibnamefont{Hoefer}}, \bibinfo{journal}{Phys. Rev. B}
  \textbf{\bibinfo{volume}{89}}, \bibinfo{pages}{180409}
  (\bibinfo{year}{2014}).

\bibitem[{\citenamefont{Maci\`a et~al.}(2012)\citenamefont{Maci\`a, Warnicke,
  Bedau, Im, Fischer, and Kent}}]{Macia:jmmm2012}
\bibinfo{author}{\bibfnamefont{F.}~\bibnamefont{Maci\`a}},
  \bibinfo{author}{\bibfnamefont{P.}~\bibnamefont{Warnicke}},
  \bibinfo{author}{\bibfnamefont{D.}~\bibnamefont{Bedau}},
  \bibinfo{author}{\bibfnamefont{M.-Y.} \bibnamefont{Im}},
  \bibinfo{author}{\bibfnamefont{D.~A.} \bibnamefont{Fischer},
  \bibfnamefont{P.~Arena}}, \bibnamefont{and}
  \bibinfo{author}{\bibfnamefont{A.~D.} \bibnamefont{Kent}},
  \bibinfo{journal}{Jour. of Mag. Mag. Mat.} \textbf{\bibinfo{volume}{324}},
  \bibinfo{pages}{3632} (\bibinfo{year}{2012}).

\bibitem[{\citenamefont{Rippard et~al.}(2010)\citenamefont{Rippard, Deac,
  Pufall, Shaw, Keller, Russek, Bauer, and Serpico}}]{Rippard2010}
\bibinfo{author}{\bibfnamefont{W.~H.} \bibnamefont{Rippard}},
  \bibinfo{author}{\bibfnamefont{A.~M.} \bibnamefont{Deac}},
  \bibinfo{author}{\bibfnamefont{M.~R.} \bibnamefont{Pufall}},
  \bibinfo{author}{\bibfnamefont{J.~M.} \bibnamefont{Shaw}},
  \bibinfo{author}{\bibfnamefont{M.~W.} \bibnamefont{Keller}},
  \bibinfo{author}{\bibfnamefont{S.~E.} \bibnamefont{Russek}},
  \bibinfo{author}{\bibfnamefont{G.~E.~W.} \bibnamefont{Bauer}},
  \bibnamefont{and} \bibinfo{author}{\bibfnamefont{C.}~\bibnamefont{Serpico}},
  \bibinfo{journal}{Phys. Rev. B} \textbf{\bibinfo{volume}{81}},
  \bibinfo{pages}{014426} (\bibinfo{year}{2010}).

\bibitem[{\citenamefont{Mohseni et~al.}(2011)\citenamefont{Mohseni, Sani,
  Persson, Anh~Nguyen, Chung, Pogoryelov, and Akerman}}]{akerman_pss2011}
\bibinfo{author}{\bibfnamefont{S.~M.} \bibnamefont{Mohseni}},
  \bibinfo{author}{\bibfnamefont{S.~R.} \bibnamefont{Sani}},
  \bibinfo{author}{\bibfnamefont{J.}~\bibnamefont{Persson}},
  \bibinfo{author}{\bibfnamefont{T.~N.} \bibnamefont{Anh~Nguyen}},
  \bibinfo{author}{\bibfnamefont{S.}~\bibnamefont{Chung}},
  \bibinfo{author}{\bibfnamefont{Y.}~\bibnamefont{Pogoryelov}},
  \bibnamefont{and} \bibinfo{author}{\bibfnamefont{J.}~\bibnamefont{Akerman}},
  \bibinfo{journal}{Phys. Status Solidi RRL} \textbf{\bibinfo{volume}{5}},
  \bibinfo{pages}{432} (\bibinfo{year}{2011}).

\bibitem[{\citenamefont{Mangin and \emph{et al}}(2006)}]{mangin}
\bibinfo{author}{\bibfnamefont{S.}~\bibnamefont{Mangin}} \bibnamefont{and}
  \bibinfo{author}{\bibnamefont{\emph{et al}}}, \bibinfo{journal}{Nature
  Materials} \textbf{\bibinfo{volume}{5}}, \bibinfo{pages}{210}
  (\bibinfo{year}{2006}).

\bibitem[{\citenamefont{Ozyilmaz et~al.}(2003)\citenamefont{Ozyilmaz, Kent, and
  \emph{et al}}}]{Kent}
\bibinfo{author}{\bibfnamefont{B.}~\bibnamefont{Ozyilmaz}},
  \bibinfo{author}{\bibfnamefont{A.~D.~e.} \bibnamefont{Kent}},
  \bibnamefont{and} \bibinfo{author}{\bibnamefont{\emph{et al}}},
  \bibinfo{journal}{Phys. Rev. Lett.} \textbf{\bibinfo{volume}{91}},
  \bibinfo{pages}{067203} (\bibinfo{year}{2003}).

\bibitem[{\citenamefont{Ralph and Stiles}(2007)}]{stiles}
\bibinfo{author}{\bibfnamefont{D.~C.} \bibnamefont{Ralph}} \bibnamefont{and}
  \bibinfo{author}{\bibfnamefont{M.~D.} \bibnamefont{Stiles}},
  \bibinfo{journal}{Journal of Magnetism and Magnetic Materials}
  \textbf{\bibinfo{volume}{320}}, \bibinfo{pages}{1190} (\bibinfo{year}{2007}).

\end{thebibliography}

\end{document}